\documentclass[runningheads]{llncs}

\usepackage[T1]{fontenc}

\usepackage[hidelinks]{hyperref} 
\usepackage{graphicx}

\usepackage{color}

\mathcode`:="003A
\usepackage{amssymb}
\usepackage{amsmath}

\usepackage{diagrams}

\usepackage{tikz}
\usetikzlibrary{arrows}
\usetikzlibrary{fadings}

\usepackage{tikz-cd}
\tikzcdset{every label/.append style = {font = \small}}

\usepackage{xspace}
\def\ie{i.e.\xspace}
\def\eg{e.g.\xspace} 

\newcommand{\id}{\mathsf{id}}

\DeclareFontFamily{OT1}{pzc}{}
\DeclareFontShape{OT1}{pzc}{m}{it}{<->s*[1.30]pzcmi7t}{}
\DeclareMathAlphabet{\mathpzc}{OT1}{pzc}{m}{it}
\def\ct#1{\mathpzc{#1}} 

\newcommand{\RQ}{\mathbb{R}}
\newcommand{\NQ}{\mathbb{N}}
\newcommand{\BQ}{\Sigma}

\newcommand{\qle}{\sqsubseteq} 
\newcommand{\qlt}{\sqsubset} 
\newcommand{\qJ}{\sqcup} 
\newcommand{\qj}{\sqcup} 
\newcommand{\qM}{\sqcap}
\newcommand{\qm}{\sqcap}
\newcommand{\qT}{\otimes} 
\newcommand{\qt}{\top} 
\newcommand{\qb}{\bot} 
\newcommand{\qI}{\mathsf{u}} 
\newcommand{\qlr}{\backslash} 
\newcommand{\qrr}{/} 

\newcommand{\qwb}{\twoheaddownarrow}

\newcommand{\V}{\ct{V}}
\newcommand{\A}{\ct{A}}
\newcommand{\B}{\ct{B}}
\newcommand{\Set}{\ct{Set}}
\newcommand{\Po}{\ct{Po}}
\newcommand{\Pos}{\Po_0}
\newcommand{\MS}{\ct{Met}}
\newcommand{\Quan}{\ct{Qnt}} 
\newcommand{\Top}{\ct{Top}}

\newcommand{\s}[1]{{#1}_0}

\newcommand{\Mon}{\mathsf{Mon}}  

\newcommand{\R}{R}
\renewcommand{\r}{r}
\newcommand{\sr}{s}
\newcommand{\T}{\mathsf{T}}
\newcommand{\inT}{\mathsf{in}} 

\newcommand{\cl}[1]{\overline{#1}}
\newcommand{\cle}[1]{\widetilde{#1}}
\newcommand{\inv}[1]{{#1}^{-1}}

\newcommand{\CS}{\mathsf{C}}
\newcommand{\PS}{\mathsf{P}}
\newcommand{\setf}[1]{\ensuremath{\{{#1}\}}}
\newcommand{\setb}[2]{\ensuremath{\{{#1}\mid{#2}\}}}
\newcommand{\famb}[2]{\ensuremath{({#1}\mid{#2})}}

\newcommand{\PM}{\mathsf{P}}

\newcommand{\defeq}{\stackrel{\vartriangle}{=}}
\newcommand{\defiff}{\stackrel{\vartriangle}{\iff}}

\newcommand{\eqc}[2][]{[#2]_{#1}}

\usepackage[text={6in,9.5in},centering]{geometry}

\begin{document}

\title{Robustness in Metric Spaces over Continuous Quantales and the Hausdorff-Smyth Monad}

\titlerunning{Robustness in Metric Spaces over Continuous Quantales}

\author{Francesco Dagnino\inst{1} \and
Amin Farjudian\inst{2} \and
Eugenio Moggi\inst{1}}

\authorrunning{F. Dagnino et al.}

\institute{DIBRIS, Universit\`{a} di Genova, Italy,
\email{\{francesco.dagnino,moggi\}@unige.it}
\and
School of Mathematics, University of Birmingham, United Kingdom,
\email{A.Farjudian@bham.ac.uk}}

\maketitle              

\begin{abstract}
  Generalized metric spaces are obtained by weakening the requirements
  ({\eg}, symmetry) on the distance function and by allowing
  it to take values in structures ({\eg}, quantales) that are more
  general than the set of non-negative real numbers. Quantale-valued
  metric spaces have gained prominence due to their use
  in quantitative reasoning on programs/systems, and for
  defining various notions of behavioral metrics.

  We investigate imprecision and robustness in the
  framework of quantale-valued metric spaces, when the quantale is
  continuous. In particular, we study the relation between the
  robust topology, which captures robustness of analyses, and the
  Hausdorff-Smyth hemi-metric. To this end, we define a
  preorder-enriched monad $\PM_S$, called the Hausdorff-Smyth monad,
  and when $Q$ is a continuous quantale and $X$ is a $Q$-metric space,
  we relate the topology induced by the metric on $\PM_S(X)$ with the
  robust topology on the powerset $\PS(X)$ defined in terms of the
  metric on $X$.

\end{abstract}

\section*{Introduction}

In the 1970s, Lawvere~\cite{lawvere1973metric} proposed viewing metric
spaces as small categories enriched over the monoidal category
$\RQ_+$, whose objects are the extended non-negative real numbers,
where there is an arrow $x\to y$ if and only if $x\geq y$, and $+$ and
$0$ provide the monoidal structure.  In this way, one recovers most
notions and results about metric spaces as instances of those about
enriched categories~\cite{kelly1982basic}.

Enrichment over arbitrary monoidal categories, however, is
unnecessarily general for studying metric phenomena. Indeed, the base
of enrichment for Lawvere's metric spaces belongs to the class of
small (co)complete posetal categories, where the tensor commutes with
colimits.
These categories are called \emph{quantales} and small categories
enriched over a quantale $Q$ are dubbed \emph{$Q$-metric spaces}.
Quantales are a useful compromise between arbitrary monoidal
categories and the specific case of
$\RQ_+$~\cite{FlaggK97,HofmannST14,CookW21}.
Beside a substantial simplification of the theory, restricting to
quantales allows to use well-known order-theoretic notions
which do not have obvious counterparts in arbitrary monoidal
categories, but are crucial to relating $Q$-metric spaces to other
structures such as topological spaces.

Quantale-valued metric spaces are also increasingly used for
\emph{quantitative reasoning} on programs/systems, and for
defining various notions of \emph{behavioral metrics}
\cite{Gavazzo18,BonchiKP18,DalLagoGY19,Pistone21,SprungerKDH21,GavazzoF23}.
The use of quantitative methods is important in coping with the
uncertainty/imprecision that arises in the analysis of, {\eg},
probabilistic programs or systems interacting with physical processes.
In these contexts, quantales provide a flexible framework which allows
choosing the most suitable notion of distance for the specific
analysis one is interested in.

Quantales arise naturally also in analysis of algorithms,
  namely, \emph{costs} are values in certain quantales (see
  Example~\ref{ex:cost-quantale}), but researchers in this area
  usually consider only subsets of these quantales and their partial
  order.


  \paragraph{Motivations.}
  the notions of imprecision and robustness are relevant in the
  context of software tools for the analysis of hybrid/continuous
  systems.  These tools manipulate (formal descriptions of)
  mathematical models.
  A mathematical model is usually a simplified description of the
  system (and its environment), with the requirement that the
  simplification should be \emph{safe}, \ie, if the analysis says that
  the model satisfies a property, then the system also satisfies that
  property. Usually, safe simplification is achieved by injecting
  \emph{non-determinism} in the model (non-determinism is useful also
  to model \emph{known unknowns} in the environment and \emph{don't
    care} in the model).
  For hybrid/continuous systems there is another issue:
  \emph{imprecision} in observations.  In fact, predictions based on a
  mathematical model and observations on a real system can be compared
  only up to the precision of measurements on the
  real system.
  We say that an analysis is \emph{robust} when it can cope with
  \emph{small amounts} of imprecision in the model, \ie, if a robust
  analysis says that a model $M$ has a property, then it says so also
  for models that have a bit more non-determinism than $M$.
  Working with metric spaces makes it possible to define imprecision
  formally and to quantify the amount of non-determinism added to a
  model.
 
Following \cite{MoggiFT-ICTCS-2019}, given a metric space $X$, we can
identify analyses with monotonic maps on the complete lattice $\PS(X)$
of subsets of $X$ ordered by reverse inclusion.\footnote{The category
  of complete lattices and monotonic maps is the framework proposed in
  \cite{cousot1992abstract} for abstract interpretations.}
However, even when imprecision is made arbitrarily small, two subsets
with the same closure are \emph{indistinguishable}.  Therefore,
analyses should be considered over the complete lattice $\CS(X)$ of
closed subsets, rather than that of arbitrary subsets, and should cope
with \emph{small amounts} of imprecision in the input.  Formally, this
property was defined as continuity with respect to the \emph{robust
topology}~\cite[Def.~A.1]{Moggi_Farjudian_Duracz_Taha:Reachability_Hybrid:2018}
on $\CS(X)$.
This yields a functor from metric spaces to $T_0$-topological
spaces, which maps a metric on $X$ to the robust topology on $\CS(X)$.
An anonymous referee
suggested that the robust topology might be related to the
\emph{Hausdorff-Smyth hemi-metric} in \cite[Proposition
  1]{goubault2008simulation}, and thus the functor from metric spaces
to topological spaces might be replaced with an endofunctor on
\emph{hemi-metric spaces} (aka, Lawvere's metric spaces).

\paragraph{Contributions.}

This paper studies the link between the robust topology and the
Hausdorff-Smyth hemi-metric---as suggested by an anonymous referee
of~\cite{Farjudian_Moggi:Robustness_Scott_Continuity_Computability:2023}---and
in doing so, addresses also more general issues, namely:

\begin{enumerate}
\item The notion of \emph{imprecision} and the definition of
  \emph{robust topology} are generalized to $Q$-metric spaces when $Q$
  is a \emph{continuous quantale}, and the results
  in~\cite{MoggiFT-ICTCS-2019} are extended to this wider setting (see
  Section~\ref{sec:part1}).
  
\item \emph{Indistinguishability} is investigated in the context of
  $\Po$-enriched categories\footnote{$\Po$ denotes the category of
    preorders and monotonic maps.}  and the notion of \emph{separated
    object} is introduced. In Section~\ref{sec:Po-enriched}, we prove
  that, under certain conditions, every $\Po$-enriched monad can be
  \emph{transformed} into one that factors through the full
  sub-category of separated objects.  The conditions that allow this
  transformation hold in many $\Po$-enriched categories, such as that
  of $Q$-metric spaces and that of topological spaces.

\item The \emph{Hausdorff-Smyth $\Po$-enriched monad} $\PM_S$ is
  defined on the category of $Q$-metric spaces, with $Q$ an arbitrary
  quantale (see Section~\ref{sec:Monads}).  When $Q$ is a continuous
  quantale, the topology induced by the metric on $\PM_S(X)$ is shown
  to coincide with a topology on $\PS(X)$, called \emph{*-robust},
  defined in terms of the metric on $X$.  In general, the *-robust
  topology is included in the robust topology, but they coincide when
  $Q$ is \emph{linear and non-trivial} (\eg, $\RQ_+$).
\end{enumerate}

Although we apply the construction in Section~\ref{sec:Po-enriched}
only to the monad defined in Section~\ref{sec:Monads}, it is
applicable to other monads definable on $Q$-metric spaces (see
Section~\ref{sec:Concluding_remarks}) or on other $\Po$-enriched
categories.


\paragraph{Summary.}

The rest of the paper is organized as follows:
\begin{itemize}
\item
Section~\ref{sec:maths} contains the basic notation
and mathematical preliminaries.
\item
Section~\ref{sec:quantales} introduces
the category $\Quan$ of quantales and lax-monoidal maps, and states
some properties of continuous quantales.
\item
Section~\ref{sec:QMS} defines the $\Po$-enriched category $\MS_Q$ of
$Q$-metric spaces and short maps for a quantale $Q$, and gives some
of its properties.
\item
Section~\ref{sec:topology} introduces two topologies associated with a
$Q$-metric space when $Q$ is continuous, and characterizes the open
and closed subsets.

\item
Section~\ref{sec:Po-enriched} defines separated objects in a
$\Po$-enriched category $\A$, and shows that, under certain
assumptions on $\A$ satisfied by $\MS_Q$, every $\Po$-enriched monad
on $\A$ can be transformed (in an optimal way) into one that factors
through the full sub-category of separated objects.

\item
Section~\ref{sec:Monads} defines the Hausdorff-Smyth distance~$d_S$
and a related $\Po$-enriched monad on $\MS_Q$, characterizes the
preorder induced by $d_S$ and, when $Q$ is continuous, also the
topology induced by $d_S$.
\item
Section~\ref{sec:Concluding_remarks} contains an overview of related
work and some concluding remarks.

\item Omitted proofs appear in Appendix~\ref{appendix:proofs}.
\end{itemize}

\section{Mathematical Preliminaries}
\label{sec:maths}

In this section, we present the basic mathematical notation used
throughout the paper.  We assume basic familiarity with order
theory~\cite{Goubault-Larrecq:Non_Hausdorff_topology:2013}. We write
$\qJ S$ to denote the join (aka lub) of a set $S$, and write $\qM S$
to denote the meet (aka glb) of $S$. Binary join and meet of two
elements $x$ and $y$ are written as $x\qj y$ and $x \qm y$,
respectively. We write $\qb$ and $\qt$ to denote the bottom and top
element of a partial order $Q$, respectively, when they exist.

We also assume basic familiarity with category
theory~\cite{borceux1994handbook}. In this article:
\begin{itemize}
\item $\Set$ denotes the category of sets and functions (alias maps).
\item $\Po$ denotes the category of preorders and monotonic maps.
\item $\Pos$ denotes the full (reflective) sub-category of $\Po$
    consisting of posets.
  \item $\Top$ denotes the category of topological spaces and
    continuous maps.
\item $\Top_0$ denotes the full (reflective) sub-category of $\Top$
  consisting of $T_0$-spaces.
\end{itemize}
All categories above have small limits and colimits.  $\Set$, $\Po$
and $\Pos$ have also exponentials, thus they are examples of
\emph{symmetric monoidal closed} categories~\cite{kelly1982basic}.
$\Po$ and $\Top$ (and their sub-categories) can be viewed as
\emph{$\Po$-enriched} categories~\cite{kelly1982basic}, \eg, the
hom-set $\Po(X,Y)$ of monotonic maps from $X$ to $Y$ can be equipped
with the pointwise preorder induced by the preorder $Y$.

Other categories introduced in subsequent sections are $\Po$-enriched,
and this additional structure is relevant when defining adjunctions
and equivalences between two objects of a $\Po$-enriched category.

\begin{definition}[Adjunction] \label{def:po-adj} Given a pair of maps
  \begin{tikzcd}[column sep = large]
    X \arrow[r, "f" description , yshift =
    1.1ex] & Y  \arrow[l, "g" description , yshift = -1.1ex]
  \end{tikzcd}in a $\Po$-enriched category $\A$, we say that they
  form:
\begin{enumerate}
\item an \textbf{adjunction} (notation $f\dashv g$) $\defiff$
  $f\circ g\leq\id_Y$ and $\id_X\leq g\circ f$, in which $f$ and $g$
  are called left- and right-adjoint, respectively.
\item an \textbf{equivalence} $\defiff$ $\id_Y\leq f\circ
  g\leq\id_Y$ and $\id_X\leq g\circ f\leq \id_X$.
\end{enumerate}
\end{definition}

We use `$\in$' for set membership ({\eg}, $x \in X$), but we use `$:$'
for membership of function types (\eg, $f: X \to Y$) and to denote
objects and arrows in categories (\eg, $X:\Top$ and $f:\Top(X,Y)$).
The powerset of a set $X$ is denoted by $\PS(X)$. Subset inclusion is
denoted by $\subseteq$, whereas strict (proper) subset inclusion is
denoted by $\subset$.  The finite powerset ({\ie}, the set of finite
subsets) of $X$ is denoted by $\PS_f(X)$, and $A\subseteq_f B$ denotes
that $A$ is a finite subset of $B$.

We denote with $\omega$ the set of natural numbers, and identify a
natural number with the set of its predecessors, {\ie}, $0 =
\emptyset$ and $n = \setf{0, \ldots, n-1}$, for any $n \geq 1$.

\section{Quantales}
\label{sec:quantales}

Conceptually, a
quantale~\cite{Mulvey:Second_topology_conference:1986,Niefield_Rosenthal:quantales:1988,AbramskyV93}
is a degenerate case of monoidal category~\cite{kelly1982basic}, in
the same way that a partial order is a degenerate case of category.

\begin{definition}[Quantale]
  \label{def:quantale}
  A \textbf{quantale} $(Q,\qle,\qT)$ is a complete lattice $(Q,\qle)$
  with a monoid structure $(Q,\qT,\qI)$ satisfying the following
  distributive laws:
  \begin{equation*}
        x\qT (\qJ S) = \qJ \setb{x\qT y}{y\in S}\quad
        \text{ and } \quad
        (\qJ S)\qT x = \qJ \setb{y\qT x}{y\in S},
  \end{equation*}
  for any $x\in Q$ and $S\subseteq Q$.
  A quantale is \textbf{trivial} when $\bot=\qI$ (which implies that
  $\forall x \in Q.\ \bot=x$), \textbf{affine} when $\qI = \qt$,
  \textbf{linear} when $\qle$ is a linear order, and
  \textbf{commutative} when $\qT$ is commutative (in this case the
  two distributive laws are inter-derivable).  A
  \textbf{frame}\footnote{Alternative names for frame are locale and
  Heyting algebra, see \cite{Johnstone86}.} is a
  quantale where $\qT=\qm$ (thus, necessarily commutative and affine).
\end{definition}

The complete lattice $(Q,\qle)$ amounts to a complete and cocomplete
category, while the distributivity laws imply that:

\begin{itemize}
\item $\qT$ is monotonic. Thus, $(Q,\qT,\qI)$ makes $(Q,\qle)$ a
  (strict) monoidal category.
\item $\qT$ (viewed as a functor) preserves colimits, in particular
  $\qb\qT x=\qb=x\qT\qb$.
\end{itemize}
\noindent
These properties imply that the functors $x\qT -$ and $-\qT
y$, have right-adjoints $x\qlr -$ and $-\qrr y$, \ie, $x\qT y\qle z
\iff y\qle x\qlr z$ and $x\qT y\qle z \iff x\qle z\qrr y$, called
left- and right-residual, respectively.
In commutative quantales (\ie, degenerate examples of symmetric
monoidal closed categories) $x\qlr z=z\qrr x$ is denoted as $[x,z]$
and is given by $[x,z] = \qJ\setb{y}{x\qT y\qle z}$.
\begin{example}\label{ex:quantale}
  We present some examples of quantales.  The first four examples
  describe linear, commutative and affine quantales (some are
  frames).  The last two items (excepts in degenerate cases) give
  non-linear, non-commutative and non-affine quantale.  The
  construction $Q/\qI$ always returns an affine quantale and preserves
  the linearity and commutative properties, while $\prod_{j \in J}Q_j$
  and $Q^P$ preserve the affine and commutative properties.
\begin{enumerate}
\item The quantale $\RQ_+$ of~\cite{lawvere1973metric} is
  the set of non-negative real numbers extended with $\infty$, with
  $x\qle y\defiff x\geq y$ and $x\qT y\defeq x+y$.  Therefore, $\qJ
  S=\inf S$, $\qM S=\sup S$, $\qb=\infty$, $\qI=\qt=0$, $[x,z]=z-x$ if
  $x\leq z$ else $0$.
  
\item $\RQ_\qm$ is similar to $\RQ_+$, but
  $x\qT y\defeq x\qm y=\max(x,y)$.
  Thus, $\RQ_\qm$ is a frame, $\qI=0$, $[x,z]=z$ if $x\leq z$ else
  $0$ ($\qt$, $\qb$, $\qJ S$, and $\qM S$ are the same as in $\RQ_+$).
  
\item $\NQ_+$ is the \emph{sub-quantale} of $\RQ_+$ whose carrier is
  the set of natural numbers extended with $\infty$.  $\NQ_\qm$ is
  the \emph{sub-frame} of $\RQ_\qm$ with the same carrier as
  $\NQ_+$.
  
\item $\BQ$ is the \emph{sub-quantale} of $\RQ_+$ whose carrier is
  $\setf{0,\infty}$. $\BQ$ is a frame.

\item $Q/\qI$ is the \emph{sub-quantale} of $Q$ whose carrier is
  $\setb{x \in Q}{x\qle\qI}$. Thus, $\qI$ is the top element of
  $Q/\qI$.

\item
  $\prod_{j \in J}Q_j$ is the product of the quantales $Q_j$, with
  $\qle$ and $\qT$ defined pointwise.
\item
  $Q^P$ is the quantale of monotonic maps from the poset $P$ to
  the quantale $Q$, with $\qle$ and $\qT$ defined pointwise.

\item
  $(\PS(M),\subseteq,\qT)$ is the quantale (actually a boolean
  algebra) of subsets of the monoid $(M,\cdot,e)$, with
  $\qI=\setf{e}$ and $A\qT B\defeq\setb{a\cdot b}{a\in A, b\in B}$.

\item $(\PS(X^2),\subseteq,\qT)$ is the quantale (boolean algebra) of
  relations on the set $X$, with $\qI=\setb{(x,x)}{x\in X}$ and:
  \begin{equation*}
  R\qT
  S\defeq\setb{(x,z)}{\exists y \in X.(x,y)\in R, (y,z)\in S}.  
  \end{equation*}
\end{enumerate}
\end{example}

\begin{example}\label{ex:cost-quantale}
  We consider some quantales arising in the analysis of algorithms.  We
  identify algorithms with multi-tape deterministic Turing Machines
  (TM), which accept/reject strings written in a finite input alphabet
  $A$.  In this context, one is interested in quantale-valued cost
  functions $X\to Q$, rather than distances.
  \begin{itemize}
    \item The size $s(w)$ of an input $w$ for a TM is a value in the
      quantale $\NQ_+$, namely the length of the string $w$.
      In particular, the size of an infinite string is $\infty$, and
      the size of the concatenation of two strings is the sum of their
      sizes.

    \item The time (\ie, the number of steps) taken by a TM on a
      specific input $w$ is again a value in $\NQ_+$.
      In particular, a TM failing to terminate on $w$ takes time
      $\infty$, and the time taken for executing sequentially two TMs
      on $w$ is the sum of the times taken by each TM (plus a linear
      overhead for copying $w$ on two separate tapes, so that the two
      TMs work on disjoint sets of tapes).
  \end{itemize} 
  The time complexity associated to a TM typically depends on the
  input (or its size), thus it cannot be a cost in $\NQ_+$.  Such cost
  should be drawn from a quantale reflecting this dependency, namely a
  \emph{higher-order} quantale.\footnote{This resembles higher-order
    distances used to compare functional programs
    \cite{DalLagoGY19,Pistone21}.} We now describe some of such
  quantales from the most precise to the most abstract.
  \begin{enumerate} 
  \item The most precise quantale is $\NQ_+^{A^*}$ (\ie, the product
    of $A^*$ copies of $\NQ_+$).  A $t\in\NQ_+^{A^*}$ maps each
    finite input $w\in A^*$ to the time taken by a TM on $w$.
  
  \item A first abstraction is to replace $t\in\NQ_+^{A^*}$ with
    $T\in\NQ_+^\omega$, where $T(n)$ is the best upper-bound for the
    time taken by a TM on inputs of size $n$, \ie,
    $T(n)=\max\setb{t(w)}{s(w)=n}$.

  \item In practice (by the linear speed-up theorem), time
    complexity is given in $O$-notation, \ie, $T\in\NQ_+^\omega$ is
    replaced with the subset $O(T)$ of $\NQ_+^\omega$ such that
    $T'\in O(T)\iff\forall n\geq n_0.T'(n)\leq C*T(n)\mbox{ for
    some $n_0$ and $C$ in $\omega$.}$

  If we replace $\NQ_+^\omega$ with the partial order $L_O$ of
  $O$-classes $O(T)$ ordered by reverse inclusion, we
  get a distributive lattice (\ie, binary meets distribute over finite
  joins, and conversely): the top is $O(0)$, the bottom is
  $O(\infty)$, the join $O(T_1)\qj O(T_2)$ is $O(T_1)\cap
  O(T_2)=O(T_1\qj T_2)=O(\min(T_1,T_2))$, the meet $O(T_1)\qm O(T_2)$
  is $O(T_1\qm T_2) = O(\max(T_1,T_2)) = O(T_1+T_2)$.

  The lattice $L_O$ is distributive, because the complete lattice
  underlying $\NQ_+^\omega$ is distributive, but it is not a frame
  (as it fails to have arbitrary joins).
  However, there is a general construction, see \cite[page
  69]{Johnstone86}, which turns a distributive lattice $L$ into
  the \emph{free frame} $I(L)$ over $L$.  More precisely, $I(L)$ is
  the poset of \emph{ideals} in $L$ ordered by inclusion, and the
  embedding $x\mapsto\downarrow x$ from $L$ to $I(L)$ preserves finite
  meets and joins.

\item A simpler way to obtain a frame is to take the subset of $L_O$
  consisting of the $O(n^k)$ with $k\in [0,\infty]$.  This linear
  frame is isomorphic to $\NQ_\qm$, namely $k\in\NQ_\qm$
  corresponds to $O(n^k)$.
  \end{enumerate}
\end{example}

There are several notions of morphism between quantales, we consider
those corresponding to lax and strict monoidal functors.
\begin{definition}\label{def:qmaps}
  A monotonic map $h:Q\to Q'$ between quantales is called:
  \begin{itemize}
  \item \textbf{lax-monoidal} $\defiff$ $\qI'\qle'h(\qI)$ and
    $\forall x,y \in Q.h(x)\qT'h(y)\qle h(x\qT y)$;
  \item \textbf{strict-monoidal} $\defiff$
        $\qI'=h(\qI)$ and $\forall x,y \in Q.h(x)\qT'h(y)=h(x\qT y)$.
  \end{itemize}
  $\Quan$ denotes the $\Pos$-enriched category of quantales and
  lax-monoidal maps, where $\Quan(Q,Q')$ has the pointwise order
  induced by the order on $Q'$.
\end{definition}
We give some examples of monotonic maps between quantales.
\begin{example}\label{ex:monoidal-maps}
  In the following diagram we write%
\begin{tikzcd}
  {} \arrow[r, dashed] &  {}
\end{tikzcd}for lax- and%
\begin{tikzcd}
  {} \arrow[r] &  {}
\end{tikzcd}for strict-monoidal maps, $1$ for the trivial quantale
(with only one element $*$), $!_Q$ for the unique map from $Q$ to
$1$, and $f\dashv g$ for ``$f$ is left-adjoint to $g$'':
\begin{equation*}
  \begin{tikzcd}[row sep = large, column sep = large]
    1 \arrow[r, dashed, "\qt_Q", "\top"', yshift = 1.1ex] & Q
    \arrow[l, "!_Q", yshift = -1.1ex] \arrow[r, dashed, "g", "\top"', yshift =
    1.1ex] & Q/\qI \arrow[l, "f", yshift = -1.1ex] \arrow[r, "g'", "\top"',
    yshift = 1.1ex] & \Sigma \arrow[l, "f'", yshift = -1.1ex] 
  \end{tikzcd} \quad
  \begin{tikzcd}[row sep = large, column sep = large]
    \NQ_+ \arrow[r, "i"', "\top", yshift = -1.1ex] & \RQ_+
    \arrow[l, dashed, "c"', yshift = 1.1ex]  & \RQ_\qm \arrow[l, dashed, "id"']
    \end{tikzcd}
\end{equation*}
\begin{itemize}
\item $\qt_Q$ maps $*$ to $\qt$;
\item $f$ is the inclusion of $Q/\qI$ into $Q$, and $g$ maps $x$ to
  $x\qm\qI$;
\item $f'$ maps $\qb$ to $\qb$ and $\qt$ to $\qt$, and
  $g'$ maps $\qt$ to $\qt$ and $x\qlt\qt$ to $\qb$;
\item $i$ is the inclusion, $c(x)=\lceil x\rceil$ is integer round up,
  and $id$ is the identity.
\end{itemize}
The frames for measuring the time complexity of TMs (see
Example~\ref{ex:cost-quantale}) are related by obvious monoidal maps
going from the more precise to the more abstract frame:
  \begin{equation*}
  \begin{tikzcd}[row sep = large, column sep = large]
    \NQ_+^{A^*} \arrow[r,dashed,"f"] &
    \NQ_+^\omega \arrow[r,"g"] &
    I(L_O) \arrow[r,dashed,"h"] & \NQ_\qm
    \end{tikzcd}
\end{equation*}
\begin{itemize}
\item $f$ maps $t\in\NQ_+^{A^*}$ to $T\in\NQ_+^\omega$ such that
  $T(n)=\max\setb{t(w)}{s(w)=n}$;
\item $g$ maps $T\in\NQ_+^\omega$ to the principal ideal
  $\downarrow O(T)\in I(L_O)$;
\item $h$ maps $X\in I(L_O)$ to $n\in\NQ_\qm$ such that
  $n=\min\setb{k}{\forall A\in X.A\subseteq O(n^k)}$.
\end{itemize}
\end{example}

\subsection{Continuous Quantales}

To reinterpret in quantale-valued metric spaces the common
$\epsilon$-$\delta$ definition of continuous maps, and relate such
spaces to topological spaces, we restrict to \emph{continuous
  quantales}, \ie, quantales whose underlying lattices are continuous.
Note that linear quantales are always continuous.
We recall the definition of a continuous lattice and related notions.
More details may be found
in~\cite{Compendium:Book:1980,AbramskyJung94-DT,Goubault-Larrecq:Non_Hausdorff_topology:2013}.

\begin{definition}
\label{def:way_below}
Given a complete lattice $(Q,\qle)$ and $x,y \in Q$, we say that:
\begin{enumerate}
\item $D \subseteq Q$ is directed $\defiff \forall x,y \in D. \exists
  z \in D. x \qle z$ and $y \qle z$ .
\item $x$ is \textbf{way-below} $y$ (notation $x\ll_Q y$, or $x\ll y$
  when $Q$ is clear from the context) $\defiff$ for any directed
  subset $D$ of $Q$, $y\qle\qJ D\implies\exists d \in D.x\qle d$.
\item $x$ is \textbf{compact} $\defiff x \ll x$.
\end{enumerate}
We write $\qwb y$ for $\setb{x \in Q}{x\ll y}$, and $Q_0$ for the set
of compact elements in $Q$.
\end{definition}

The following are some basic properties of the way-below relation.
\begin{proposition}\label{thm:wb:prop}
  In any complete lattice $(Q,\qle)$, and for all $x, x_0, x_1 \in Q$:
\begin{enumerate}
\item $x_0\ll x_1\implies x_0\qle x_1$.
\item $x_0'\qle x_0\ll x_1\qle x_1'\implies x_0'\ll x_1'$.
\item $\bot\ll x$.
\item $\qwb x$ is directed. In particular, $x_0,x_1\ll x\implies x_0\qj x_1\ll x$.
\end{enumerate}
\end{proposition}

\begin{definition}[Continuous Lattice]\label{def:CL}
Given a complete lattice $Q$, we say that:
\begin{enumerate}
\item $Q$ is \textbf{continuous} $\defiff\forall
  x \in Q.x=\qJ\qwb x$.
\item $B\subseteq Q$ is a \textbf{base} for $Q$ $\defiff\forall x \in X.
  B\cap\qwb x$ is directed and
  $x=\qJ(B\cap\qwb x)$.
\item $Q$ is \textbf{$\omega$-continuous} $\defiff$ $Q$ has a countable base.
\item $Q$ is \textbf{algebraic} $\defiff Q_0$ is a base for $Q$.
\end{enumerate}
\end{definition}

A complete lattice $Q$ is continuous exactly when it has a base.
Any base for $Q$ must includes $Q_0$. The set $Q_0$ is a base only
when $Q$ is algebraic and the bottom element $\bot$ is always
compact.
Continuous lattices enjoy the following interpolation property
(see~\cite[Lemma~2.2.15]{AbramskyJung94-DT}):

\begin{lemma}
  \label{lemma:inter:cl}
  For any continuous lattice $Q$ and $q_1, q_2 \in Q$, $q_1 \ll q_2\implies\exists q \in
  Q.q_1\ll q\ll q_2$.
\end{lemma}
Continuous quantales enjoy a further interpolation property:
\begin{lemma}
  \label{lemma:inter:cq}
  In every continuous quantale,
  $q_1 \ll q_2\implies\exists q\ll\qI.q_1\ll q_2\qT q$ and
  $q_1 \ll q_2\implies\exists q\ll\qI.q_1\ll q\qT q_2$.
\end{lemma}

\begin{proof}
  Appendix~\ref{proof:lemma:inter:cq}. \qed
\end{proof}

\begin{example}
The quantales in Example~\ref{ex:quantale} have the following properties:
\begin{itemize}
\item $\NQ_+$, $\NQ_\qm$, and $\BQ$ are $\omega$-algebraic. More
  precisely, all elements in these quantales are compact, and
  $x\ll y\iff x\geq y$ (or equivalently $x\qle y$).
\item $\RQ_+$ and $\RQ_\qm$ are $\omega$-continuous, \eg, the set of
  rational numbers with $\infty$ is a base, $x\ll y\iff (x=\infty\lor
  x>y)$, and $\infty$ is the only compact element.

\item $\PS(M)$ and $\PS(X^2)$ are algebraic, the sets of
  compact elements are $\PS_f(M)$ for $\PS(M)$ and $\PS_f(X^2)$ for
  $\PS(X^2)$, and $A\ll B\iff A\subseteq_f B$.
\end{itemize}
\end{example}

\noindent
Continuous lattices (and quantales) have the following closure
properties:
\begin{proposition}
  Continuous (algebraic) lattices are closed under small products.
  $\omega$-continuous lattices are closed under countable products.
\end{proposition}

\begin{proof}
  The claims follow from the fact that if $\forall j \in J.B_j$ is a
  base for $Q_j$, then
  $\setb{x \in \prod_{j \in J}B_j}{\exists J_0\subseteq_fJ.\forall j
    \in J-J_0.x_j=\qb_j}$ is a base for $\prod_{j \in J}Q_j$.  \qed
\end{proof}
We conclude by observing that linear quantales are always continuous.
\begin{proposition}
  Every linear quantale is continuous.
\end{proposition}
\begin{proof}
Use \cite[Exercise 1.7]{Compendium:Book:1980}, where linearly ordered
complete lattices are called complete chains.\qed
\end{proof}

\section{Quantale-valued Metric Spaces}
\label{sec:QMS}

In~\cite{lawvere1973metric}, Lawvere views metric spaces as
$\RQ_+$-enriched categories, and shows that several definitions and
results on metric spaces are derivable from general results on
$\V$-enriched categories, where $\V$ is a symmetric monoidal closed
category (see~\cite{kelly1982basic}).
We replace $\RQ_+$ with a quantale $Q$, and consider the
$\Po$-enriched category of $Q$-metric spaces and short maps, whose
objects are $Q$-enriched small categories and whose arrows are
$Q$-enriched functors.

\begin{definition}[$\MS_Q$]\label{def:Q-MS}
  Given a quantale $Q$, the $\Po$-enriched category $\MS_Q$ of
  \textbf{$Q$-metric spaces} and \textbf{short maps} is given by:
  \begin{description}
  \item[objects] are pairs $(X,d)$ with $d:X^2\to Q$ satisfying
    $d(x,y)\qT d(y,z)\qle d(x,z)$ and $\qI\qle d(x,x)$;
    $d$ induces on $X$ the \textbf{$d$-preorder} $x\leq_d
    y\defiff\qI\qle d(x,y)$.
  \item[arrows] in $\MS_Q((X,d),(X',d'))$ are $f:X\to X'$ satisfying
    $\forall x,y\in X.d(x,y)\qle d'(f(x),f(y))$ with
    \textbf{hom-preorder} $f\leq f'\defiff\forall x \in
    X.f(x)\leq_{d'}f'(x)$.
  \end{description}

  \noindent
  An arrow $f:\MS_Q((X,d),(X',d'))$ is said to be an \textbf{isometry} when
  $\forall x,y \in X.d(x,y)=d'(f(x),f(y))$.
\end{definition}

In comparison with the properties of a standard metric $d$, we have
that:
\begin{itemize}
\item the triangular inequality $d(x,z)\leq d(x,y)+d(y,z)$ becomes
  $d(x,y)\qT d(y,z)\qle d(x,z)$. Note that, in $\RQ_+$, the order
  $\qle$ is $\geq$, and $\qT = +$;
\item $d(x,y)=0\iff x=y$ is replaced by the weaker property
  $\qI\qle d(x,x)$, which corresponds to $d(x,x)=0$. Note that in
  $\RQ_+$, we have $\qI=0=\qt$;
\item symmetry $d(x,y)=d(y,x)$ is unusual in (enriched) category
  theory.
\end{itemize}

In the absence of symmetry, \emph{separation}, {\ie},
$d(x,y)=0\implies x=y$, should be recast as
$(d(x,y)=0\land d(y,x)=0)\implies x=y$, which in a quantale setting
becomes $(\qI\qle d(x,y)\land \qI\qle d(y,x))\implies x=y$.  The
objects with this property are exactly the $(X,d)$ such that the
preorder $\leq_d$ is a poset.
Section~\ref{sec:Po-enriched} gives a more abstract definition of
separated object in a $\Po$-enriched category.

\begin{example}
We relate $\MS_Q$ for some quantales $Q$ to more familiar categories:
\begin{enumerate}
\item $\RQ_\qm$-metric spaces generalize ultrametric spaces, {\ie},
  spaces where the metric satisfies $d(x,z)\leq\max(d(x,y),d(y,z))$.
\item $\MS_\BQ$ is (isomorphic to) the $\Po$-enriched category $\Po$
  of preorders and monotonic maps, and the \emph{separated objects} of
  $\MS_\BQ$ are the posets.
\item $\MS_1$ is the category $\Set$ of sets and functions, with the
  chaotic preorder on $\Set(X,Y)$, \ie, $f\leq g$ for every
  $f,g;\Set(X,Y)$, and the \emph{separated objects} of $\MS_1$ are the
  sets with at most one element.
\end{enumerate}
\end{example}

We summarize some properties of $\MS_Q$, which ignore the
$\Po$-enrichment, proved in~\cite{kelly1982basic} for a generic
complete and cocomplete symmetric monoidal closed category in place of
a quantale $Q$.

\begin{proposition}
  \label{prop:MetQ_has_small_sums_etc}
For any quantale $Q$, the category $\MS_Q$ has small products, small
sums, equalizers and coequalizers.
\end{proposition}

\begin{proof}
  Appendix~\ref{proof:prop:MetQ_has_small_sums_etc}. \qed
\end{proof}

Lax-monoidal maps induce $\Po$-enriched functors.
\begin{definition}
Given a lax monoidal map $h:\Quan(P,Q)$, the $\Po$-enriched functor
$h:\MS_P\to\MS_Q$ is such that $h(X,d)\defeq(X,h\circ d)$ and is the
identity on arrows.
\end{definition}

\section{Topologies on $Q$-metric spaces} 
\label{sec:topology}

When $Q$ is a continuous quantale, one can establish a relation
between $\MS_Q$ and $\Top$, thereby generalizing the open ball
topology induced by a standard metric. In general, to a $Q$-metric $d$
on $X$ one can associate at least two topologies on $X$.  When $Q$ is
$\omega$-continuous---a restriction desirable from a computational
viewpoint (see~\cite{smyth1977effectively})---convergence can be
defined in terms of sequences.

\begin{definition}
  Given a continuous quantale $Q$ and $(X,d):\MS_Q$, the \textbf{open ball}
  with center $x\in X$ and radius $\delta\ll\qI$ is
  $B(x,\delta)\defeq\setb{y \in X}{\delta\ll d(x,y)}$.
The \textbf{open ball topology} $\tau_d$ is the topology
generated by the family of open balls.

One can define also the \textbf{dual open ball}
$B^o(x,\delta)\defeq\setb{y \in X}{\delta\ll d(y,x)}$, and the
corresponding \textbf{dual open ball topology} $\tau_d^o$.
\end{definition}

When $d$ is symmetric, \ie, $d(x,y)=d(y,x)$, the two notions of open
ball agree.
In the rest of this section, we focus on open balls only, but the
results hold \emph{mutatis mutandis} also for the dual notion.
The following proposition implies that open balls form a \emph{base}
for $\tau_d$, \ie, every open in $\tau_d$ is a union of open balls.

\begin{proposition}\label{prop:ob:cq}
Open balls satisfy the following properties:
\begin{enumerate}
\item\label{ob:cq:1}
  $x\in B(x,\delta)$.
\item\label{ob:cq:2}
  $\delta\qle\delta'\implies B(x,\delta')\subseteq B(x,\delta)$.
\item\label{ob:cq:3}
  $y\in B(x,\delta)\implies\exists\delta'\ll\qI. B(y,\delta')\subseteq
  B(x,\delta)$.
\item\label{ob:cq:4}
  $y\in B(x_1,\delta_1) \cap B(x_2,\delta_2)\implies \exists
  \delta'\ll \qI.\ B(y,\delta') \subseteq B(x_1,\delta_1) \cap
  B(x_2,\delta_2)$.
\end{enumerate}
\end{proposition}

\begin{proof}
  Appendix~\ref{proof:prop:ob:cq}. \qed
\end{proof}

We show that, for continuous quantales, continuity with respect to the open
ball topology can be recast in terms of the usual \emph{epsilon-delta}
formulation:
\begin{lemma}\label{lemma:e-d:open}
If $(X,d):\MS_Q$, with $Q$ continuous, and $O\subseteq X$, then
$O\in\tau_d\iff$
\begin{equation}\label{e-d:open}
  \forall x \in O.\exists\delta\ll\qI.B(x,\delta)\subseteq O.
\end{equation}
\end{lemma}

\begin{proof}
  Appendix~\ref{proof:lemma:e-d:open}. \qed
\end{proof}

The following result characterizes the closed subsets for the topology
$\tau_d$.  Informally, the closure of a subset $A$ can be described as
the set of points \emph{from} which one can reach a point in $A$ 
within any arbitrarily small distance.

\begin{lemma}\label{lemma:e-d:closed}
  If $(X,d):\MS_Q$, with $Q$ continuous, and $A \in \PS(X)$, then the
  closure of $A$ in the topological space $(X,\tau_d)$ is given by:
  \begin{equation}\label{e-d:closed}
    \cl{A}=\setb{y \in X}{\forall\delta\ll
    \qI.\exists x \in A.\delta\ll d(y,x)}.
  \end{equation}
\end{lemma}

\begin{proof}
  To prove that $\cl{A}$ is the closure of $A$, we show that
  $z\not\in\cl{A}\iff$ exists $\delta\ll\qI$ such that $B(z,\delta)$
  and $A$ are disjoint.
  The claim follows from the equivalences:
  $z\not\in\cl{A}\iff$ $\exists\delta\ll\qI.\forall x \in A.\delta\not\ll
  d(z,x)\iff$ $\exists\delta\ll\qI.B(z,\delta)\cap A=\emptyset$. \qed
\end{proof}

\begin{theorem}\label{thm:e-d:continuity}
  Given a continuous quantale $Q_i$ and an object
  $(X_i,d_i):\MS_{Q_i}$ for each $i \in \setf{1,2}$, if
  $f:X_1\to X_2$, then $f:\Top((X_1,\tau_{d_1}),(X_2,\tau_{d_2}))\iff$
\begin{equation}\label{e-d:continuity}
  \forall x \in X_1.\forall\epsilon\ll
  \qI_2.\exists\delta\ll\qI_1.f(B(x,\delta))\subseteq
  B(f(x),\epsilon).
\end{equation}
\end{theorem}

\begin{proof}
  Appendix~\ref{proof:thm:e-d:continuity}. \qed
\end{proof}

The above characterization of continuous maps suggests a variant of
$\Top$ in which the objects are $Q$-metric spaces (for some continuous
quantale $Q$) instead of topological spaces, while the rest is
unchanged (see \cite{CookW21}):
\begin{definition}
The $\Po$-enriched category $\MS_c$ of metric spaces and continuous maps
is given by:
\begin{description}
\item[objects] are the triples $(X,d,Q)$ with $Q$ continuous quantale
  and $(X,d):\MS_Q$;

\item[arrows] in $\MS_c((X,d,Q),(X',d',Q'))$ are 
  $f:\Top((X,\tau_{d}),(X',\tau_{d'}))$, or
  equivalently $f:X\to X'$ satisfying
  $\forall x\in X.\forall\epsilon\ll\qI'.\exists\delta\ll\qI.
  f(B(x,\delta))\subseteq B(f(x),\epsilon)$.
\end{description}
\end{definition}
Similarly, one can define the sub-category $\MS_u$ of $\MS_c$ with
the same objects, but whose arrows are the \emph{uniformly continuous}
maps, \ie, $f:X\to X'$ satisfying
$\forall\epsilon\ll\qI'.\exists\delta\ll\qI.\forall x\in X.
f(B(x,\delta))\subseteq B(f(x),\epsilon)$.

\subsection{Imprecision and Robustness}
\label{sec:part1}

We extend the notions of imprecision and robustness, that
in~\cite{Moggi_Farjudian_Duracz_Taha:Reachability_Hybrid:2018,MoggiFT-ICTCS-2019}
are defined for standard metric spaces, to $Q$-metric spaces for a
continuous quantale $Q$\footnote{It is possible to relax the
assumption of continuity of $Q$ along the lines of~\cite{CookW21}.}.
Since a $Q$-metric may fail to be symmetric, we must consider the
``direction'' along which the distance is measured.  In particular,
in the presence of imprecision, two subsets are indistinguishable when
they have the same closure in the dual topology $\tau_d^o$, rather
than in the topology $\tau_d$
(Proposition~\ref{prop:spec-robust-top}).  This difference cannot be
appreciated when $d$ is symmetric, because the two topologies
coincide.

\begin{definition}
Given a $Q$-metric space $(X,d)$, with $Q$ continuous, the notions
introduced in \cite[Definition 1]{MoggiFT-ICTCS-2019} can be recast as
follows:
\begin{enumerate}
\item $B_R(A,\delta)\defeq \setb{y \in X}{\exists x \in A.\delta\ll
  d(x,y)}= \cup_{x\in A}B(x,\delta)\subseteq X$ is the set of points
  belonging to $A\subseteq X$ with precision greater than
  $\delta\ll\qI$.\footnote{The terminology used in
  \cite{MoggiFT-ICTCS-2019} is ``with imprecision less than
  $\delta$''.}

\item $A_\delta\defeq\cl{B_R(A,\delta)}^o\subseteq X$ is the
  \textbf{$\delta$-flattening} of $A\subseteq X$ with $\delta\ll\qI$,
  where $\cl{A}^o$ is the closure of $A$ in $\tau_d^o$ (see
  Lemma~\ref{lemma:e-d:closed}).
\end{enumerate}
\end{definition}

\begin{proposition}
  \label{prop:BR_A_delta_properties}
The subsets $B_R(A,\delta)$ have the following properties:
\begin{enumerate}
\item \label{prop:BR_A_delta_properties:mon} 
  $A\subseteq B_R(A,\delta)\subseteq B_R(A',\delta')$ when
  $A\subseteq A'\subseteq X$ and $\delta'\qle\delta\ll\qI$.
    
\item \label{prop:BR_A_delta_properties:tensor} 
  $B_R(B_R(A,\delta_1),\delta_2)\subseteq B_R(A,\delta)$ when
  $\delta_1,\delta_2\ll\qI$ and
  $\delta\ll\delta_1\qT\delta_2[\qle\delta_i]$.

\item \label{prop:BR_A_delta_properties:cl} 
  $\cl{A}^o=\cap_{\delta\ll\qI}B_R(A,\delta)$ for every $A\subseteq X$.

\item \label{prop:BR_A_delta_properties:cl-eq} 
  $B_R(\cl{A}^o,\delta)=B_R(A,\delta)$ for every $A\subseteq X$
  and $\delta\ll\qI$, \ie, $A$ and $\cl{A}^o$ are indistinguishable in
  the presence of imprecision.

\item \label{prop:BR_A_delta_properties:falt} 
  $B_R(A,\delta)\subseteq A_\delta\subseteq B_R(A,\delta')$ when
  $A\subseteq X$ and $\delta'\ll\delta\ll\qI$.
\end{enumerate}
\end{proposition}
\begin{proof}
  Appendix~\ref{proof:prop:BR_A_delta_properties}. \qed
\end{proof}

\begin{figure}[h]
  \centering
  \scalebox{0.35}[0.35]{\includegraphics{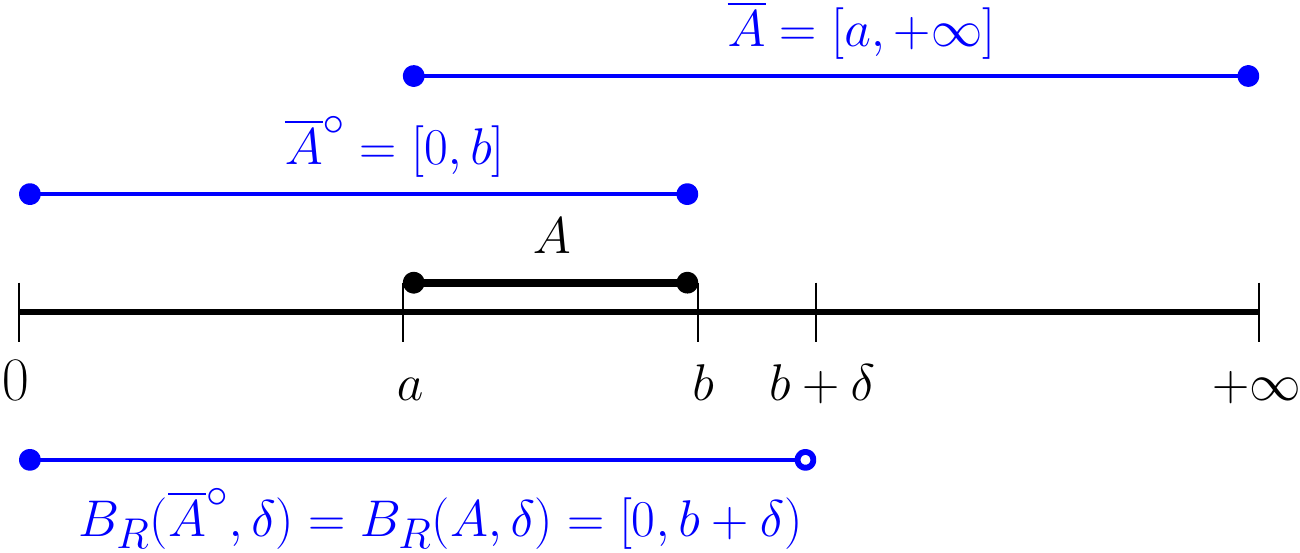}}
  \caption{Graphic recast of
    Example~\ref{example:non_symmetric_R_Plus}.}
  \label{fig:R_Plus}
\end{figure}

\begin{example}
  \label{example:non_symmetric_R_Plus}
  Consider the $Q$-metric space $(X,d)$, where $Q = X = \RQ_+$ and
  $d(x,y)\defeq\text{$y-x$ if $x \leq y$ else $0$}$.  If $A=[a,b]$ and
  $\delta \in (0,+\infty)$, then $\cl{A}= [a,+\infty]$, $\cl{A}^\circ
  = [0,b]$, and $B_R(\cl{A}^o,\delta)=B_R(A,\delta) = [0,b+\delta)$,
    as depicted in Fig.~\ref{fig:R_Plus}.
\end{example}


We can generalize to this wider setting also the definition of
robust topology in \cite[Definition
  A.1]{Moggi_Farjudian_Duracz_Taha:Reachability_Hybrid:2018}.
We define such topology on $\PS(X)$, rather than on the set of closed
subsets in the topology $\tau_d^o$, since the restriction to the set
of closed subsets amounts to replacing a topological space with an
equivalent \emph{separated} topological space (see
Section~\ref{sec:Po-enriched}).

\begin{definition}\label{def:top:UR}
  Given a $Q$-metric space $(X,d)$, with $Q$ continuous, the
  \textbf{robust topology} $\tau_{d,R}$ on $\PS(X)$ is  
  given by:
$$
  U\in\tau_{d,R}\defiff \forall A\in
  U.\exists\delta\ll\qI.\PS(B_R(A,\delta))\subseteq U.
$$ 
\end{definition}

Finally, we characterize the specialization preorder
$\leq_{\tau_{d,R}}$ induced by the robust topology $\tau_{d,R}$ on
$\PS(X)$.  As a consequence, we have that two subsets are
indistinguishable in $\tau_{d,R}$ exactly when they have the same
closure in $\tau_d^o$.

\begin{proposition}
  \label{prop:spec-robust-top}
  Let $(X,d)$ be a $Q$-metric space with $Q$ continuous, and
  $A,B\subseteq X$.  Then, we have
  $A\leq_{\tau_{d,R}} B \iff B \subseteq \cl{A}^o$.
\end{proposition}
\begin{proof}
  Appendix~\ref{proof:prop:spec-robust-top}. \qed
\end{proof}


\section{Separation in Preorder-enriched Categories}
\label{sec:Po-enriched}

Structures like preorders and topologies have a notion of
\emph{indistinguishability} between elements.  Informally, in such
structures, \emph{separation} can be understood as the property
requiring that indistinguishable elements are equal.

In this section, we define and study this notion in the setting of
$\Po$-enriched categories.  We also show that the definition
separation in this abstract setting subsumes many \emph{set-theoretic}
definitions within specific categories, in particular the category of
$Q$-metric spaces.
\begin{definition}[Separation]
  Given a $\Po$-enriched category $\A$, we say that:
  \begin{enumerate}
  \item $f,g\in\A(X,Y)$ are \textbf{equivalent} (notation
    $f\sim g$) $\defiff f\leq g\land g\leq f$.
  \item the hom-preorder $\A(X,Y)$ is separated $\defiff$ it is a
    poset.
  \item the object $Y\in\A$ is separated $\defiff\A(X,Y)$
    is separated for every $X\in\A$.
  \item $\A$ is separated $\defiff Y$ is separated for every $Y\in\A$,
    \ie, $\A$ is $\Pos$-enriched.
  \end{enumerate}
\end{definition}
\begin{remark}
  The definition of ``$\A(X,Y)$ is separated'' can be recast in terms
  of equivalence, \ie, $f\sim g\implies f=g$, for every $f,g:\A(X,Y)$.
  There is a similar recast also for the definition of ``$\A$ is
  separated'', \ie, $f\sim g\implies f=g$, for every pair $(f,g)$ of
  parallel arrows in $\A$.
  In some $\Po$-enriched categories, separated objects have a
  set-theoretic characterization that does not refer to arrows:
\begin{enumerate}
\item in $\Po$, separated objects are posets.
\item In $\Top$, separated objects are $T_0$-spaces.
\item In $\MS_Q$, separated objects are separated $Q$-metric spaces
  (see Section~\ref{sec:QMS}).
\end{enumerate}
\end{remark}

Recall from \cite{kelly1982basic} that a $\Po$-enriched functor
$F:\A\rTo\B$ is \textbf{full\&faithful} (notation $F:\A\rInto\B$) when
the maps $F_{X,Y} : \A(X,Y) \to \B(FX,FY)$ are iso in $\Po$, and a
$\Po$-enriched sub-category $\A$ of $\B$ is full when the
$\Po$-enriched inclusion functor is full\&faithful.
\begin{definition}
  \label{def:s-construction}
  If $\A$ is a $\Po$-enriched category, then $\s\A$ denotes the full
  sub-category of separated objects in $\A$.
\end{definition}

If every object in $\A$ is separated, then $\s\A$ is equal to $\A$.  A
weaker property is that every object in $\A$ is \emph{equivalent} (in
the sense of Definition~\ref{def:po-adj}) to one in $\s\A$.  This
weaker property holds in $\Po$, $\Top$, and $\MS_Q$.
\begin{proposition}
  \label{prop:MetQ_every_object_separated}
  In $\MS_Q$, every object is equivalent to a separated one.
\end{proposition}

\begin{proof}
  Appendix~\ref{proof:prop:MetQ_every_object_separated}. \qed
\end{proof}

If every object in $\A$ is equivalent to a separated one, then every
$\Po$-enriched endofunctor on $\A$ can be \emph{transformed} into one
that factors through $\s\A$.  This \emph{transformer} lifts to the
category of $\Po$-enriched monads on $\A$.

\begin{definition}\label{def:Mon}
  Given a $\Po$-enriched category $\A$, we denote by $\Mon(\A)$ the
  category of \textbf{$\Po$-enriched monads} on $\A$ and \textbf{monad
    maps}, \ie
  \begin{description}
  \item[objects:] $\Po$-enriched monads on $\A$, \ie, triples
    $\hat{M}=(M,\eta,-^*)$, where:
    \begin{itemize}
    \item $M$ is a function on the objects of
      $\A$,

    \item $\eta$ is a family of arrows $\eta_X: \A(X,MX)$ for
      $X: \A$,
    \item $-^*$ is a family of monotonic maps $\A(X,MY)\to\A(MX,MY)$
      between hom-preorders for $X,Y: \A$,
    \end{itemize}
    and satisfy the equations:
  \begin{equation}\label{eq:monad}
    \eta_X^*=\id_{MX}\quad,\quad
    f^*\circ\eta_X=f\quad,\quad
    g^*\circ f^*=(g^*\circ f)^*.
  \end{equation}
  
\item[arrows:] $\theta$ from $\hat{M}$ to $\hat{M}'$ are families of
  maps $\theta_X:\A(MX,M'X)$ for $X: \A$
  satisfying the equations:
  \begin{equation}\label{eq:monad:map}
    \theta_X\circ\eta_X=\eta'_X\quad,\quad
    \theta_Y\circ f^*=(\theta_Y\circ f)^{*'}\circ\theta_X .
  \end{equation}
  \end{description}
\noindent
  A basic \textbf{monad transformer} on $\Mon(\A)$ is a pair
  $(\T,\inT)$, where $\T$ is function on the objects of $\Mon(\A)$ and
  $\inT$ is a family of monad maps $\inT_{\hat{M}}$ from $\hat{M}$ to
  $\T\hat{M}$.
\end{definition}

\begin{remark}
  The category $\Mon(\A)$ can be made $\Po$-enriched.  The enrichment
  is relevant for defining equivalence of monads.  For our purposes,
  however, it suffices to relate (by a monad map) a generic
  $\Po$-enriched monad on $\A$ to one that factors through $\s\A$.
\end{remark}

We use the simplest form of monad transformer among those in the
taxonomy of \cite{JMoggi2010monad}, \ie, \emph{basic transformer}.
However, the monad transformer described in the following theorem can
be shown to be a \emph{monoidal transformer}.
\begin{theorem}\label{thm:endo:transformer}
  If $\A$ is a $\Po$-enriched category and $\famb{\r_X:X\to\R X}
  {X: \A}$ is a family of arrows in $\A$ such that:
  \begin{equation}\label{prop:weak:equiv}
    \text{$\R X: \s\A$ and $(\r_X,\sr_X)$ is an equivalence for some $\sr_X:\R X\to X$},
  \end{equation}
  then $(\T,\inT)$ defined below is a monad transformer on
  $\Mon(\A)$:
\begin{itemize}
\item $\T$ is the function mapping $\hat{M}=(M,\eta,-^*)$
  to $\T\hat{M}=(M',\eta',-^{*'})$, where
  \begin{itemize}
  \item $M'X\defeq \R(MX)$
  \item $\eta'_X\defeq \r_{MX}\circ\eta_X:\A(X,M'X)$
  \item if $f:\A(X,M'Y)$, then
    $f^{*'}\defeq\r_{MY}\circ(\sr_{MY}\circ f)^*\circ\sr_{MX}:\A(M'X,M'Y)$.
  \end{itemize}

\item $\inT$ is the family of monad maps such that
  $\inT_{\hat{M},X}\defeq\r_{MX}:\A(MX,M'X)$.
\end{itemize}
Moreover, the definition of $\T$ is independent of the choice of $\sr_X$.
\end{theorem}

\begin{proof}
  Appendix~\ref{proof:thm:endo:transformer}. \qed
\end{proof}

\section{The Hausdorff-Smyth Monad}
\label{sec:Monads}

In this section, we introduce a $\Po$-enriched monad $\PM_S$ on
$\MS_Q$, related to the Hausdorff-Smyth hemi-metric in
\cite{goubault2008simulation}, which \emph{extends} the powerset monad
$\PM$ on $\Set$ to $Q$-metric spaces.
By applying the monad transformer $\T$ defined in
Section~\ref{sec:Po-enriched}, one obtains a separated version of
$\PM_S$, which amounts to partitioning $\PM(X)$ into equivalence
classes, for which we define canonical representatives.
Finally, we investigate the relation between $\PM_S$ and the robust topology
in Definition~\ref{def:top:UR}.

Recall that the monad $(\PM,\eta,-^*)$ on $\Set$ is given by
$\eta_X : \Set(X,\PM(X))$ and
$-^*:\Set(X,\PM(X'))\to\Set(\PM(X),\PM(X'))$, where:
\begin{eqnarray*}
  \eta(x)&=&\{x\},\\
  f^*(A)&=&\bigcup_{x \in A}f(x).
\end{eqnarray*}

\begin{definition}[The $\PM_S$ monad]\label{def:MHS}
  Let $\PM_S$ be the function on $Q$-metric spaces such that
  $\PM_S(X,d)=(\PM(X),d_S)$, where $d_S:\PM(X)^2\to Q$ is given by:
  \begin{equation*}
    d_S(A,B)=\qM_{y \in B}\qJ_{x \in A}d(x,y).
  \end{equation*}
  The rest of the monad structure for $\PM_S$, \ie, the unit $\eta$
  and the Kleisli extension $-^*$, is inherited from that for $\PM$.
  In particular, $\eta_{(X,d)}=\eta_X$.
\end{definition}

We now prove that what we have defined is a $\Po$-enriched monad on
$\MS_Q$.

\begin{proposition}
  \label{prop:PS_monad_MetQ}
  The triple $(\PM_S,\eta,-^*)$ is a $\Po$-enriched monad on $\MS_Q$,
  \ie
  \begin{enumerate}
\item $(\PM(X),d_S):\MS_Q$, \ie, $\qI\qle d_S(A,A)$ and $d_S(A,B)\qT
  d_S(B,C)\qle d_S(A,C)$.    
  \item $\eta:\MS_Q(X,\PM_S(X))$.
  \item $f:\MS_Q(X,\PM_S(X'))$ implies $f^*:\MS_Q(\PM_S(X),\PM_S(X'))$.    
  \item $f\leq g$ in $\MS_Q(X,\PM_S(X'))$ implies $f^*\leq g^*$ in
  $\MS_Q(\PM_S(X),\PM_S(X'))$.
  \end{enumerate}
Moreover, $(\PM_S,\eta,-^*)$ satisfies the equations~(\ref{eq:monad})
for a monad.
\end{proposition}

\begin{proof}
  Appendix~\ref{proof:prop:PS_monad_MetQ}. \qed
\end{proof}

The Hausdorff-Smyth metric $d_S$ induces a preorder $\leq_{d_S}$ and
an equivalence $\sim_{d_S}$ on $\PS(X)$.  In the following, we define
the canonical representative for the equivalence class of
$A\subseteq X$ with respect to $\sim_{d_S}$, called the *-closure of
$A$, which turns out to be the biggest subset of $X$ in the
equivalence class.

\begin{definition}
Given a $Q$-metric space $(X,d)$, we define:
\begin{enumerate}
\item $d(A,y)\defeq\qJ_{x \in A}d(x,y)\in Q$ the \textbf{*-distance}
  from $A\subseteq X$ to $y\in X$.
\item $\cle{A}\defeq\setb{y \in X}{\qI\qle d(A,y)}$  the \textbf{*-closure}
  of $A\subseteq X$.
\end{enumerate}
\end{definition}
\begin{proposition}
For every $Q$-metric space $(X,d)$ the following properties hold:
\begin{enumerate}
\item $d_S(A,B)=\qM_{y \in B}d(A,y)$ and $d(A,y)=d_S(A,\setf{y})$.
\item $A\leq_{d_S} B\iff B\subseteq\cle{A}$.
\item $A\sim_{d_S} B\iff \cle{A}=\cle{B}$.
\end{enumerate}
\end{proposition}

\begin{proof}
For each property we give a proof hint.
\begin{enumerate}
\item The two equalities follow easily from the definition of $d_S$.
\item We have the following chain of equivalences:
  \begin{itemize}
  \item $A\leq_{d_S} B\iff\qI\qle d_S(A,B)\iff$
  \item $\forall y \in B.\qI\qle d(A,y)\iff$
  \item $\forall y \in B.y\in\cle{A}\iff B\subseteq\cle{A}$.
  \end{itemize}
\item Immediate by the characterization of $\leq_{d_S}$.  \qed
\end{enumerate}
\end{proof}

\subsection{Hausdorff-Smyth and *-Robust Topology}
\label{sec:part2}

We give a characterization of the topology $\tau_{d_S}$ on $\PS(X)$
using a topology $\tau_{d,S}$ defined by analogy with the robust
topology $\tau_{d,R}$ of Section~\ref{sec:part1}.
In summary, we have that $\tau_{d_S}=\tau_{d,S}\subseteq\tau_{d,R}$
when $Q$ is continuous, and $\tau_{d,S}=\tau_{d,R}$ when $Q$ is linear
and non-trivial.

\begin{definition}
Given a $Q$-metric space $(X,d)$, with $Q$ continuous, we define the
topology $\tau_{d,S}$ on $\PS(X)$:
\begin{enumerate}
\item $B_S(A,\delta)\defeq\setb{y \in X}{\delta\ll d(A,y)}\subseteq X$
  is the set of points belonging to $A\subseteq X$ with *-precision
  greater than $\delta\ll\qI$.

\item the \textbf{*-robust topology} $\tau_{d,S}$ on $\PS(X)$ is given
  by:
  
  $U\in\tau_{d,S}\defiff \forall A\in
  U.\exists\delta\ll\qI.\PS(B_S(A,\delta))\subseteq U$.
\end{enumerate}
\end{definition}

\begin{lemma}
  \label{lemma:BS_delta_properties}
The subsets $B_S(A,\delta)$ have the following properties:
\begin{enumerate}
\item $B_R(A,\delta)\subseteq B_S(A,\delta)\subseteq B_S(A',\delta')$
  when $A\subseteq A'\subseteq X$ and $\delta'\qle\delta\ll\qI$.
  
\item $\delta\qle d_S(A,B_S(A,\delta))$ for every $A\subseteq X$ and
  $\delta\ll\qI$.
\end{enumerate}
\end{lemma}

\begin{proof}
  Appendix~\ref{proof:lemma:BS_delta_properties}. \qed
\end{proof}

\begin{proposition}\label{prop:top:SR}
  For every $Q$-metric
  space $(X,d)$ with $Q$ continuous:
  \begin{equation*}
  \tau_{d_S}=\tau_{d,S}\subseteq\tau_{d,R}.    
  \end{equation*}
\end{proposition}

\begin{proof}
  Appendix~\ref{proof:prop:top:SR}. \qed
\end{proof}

\begin{lemma}\label{lem:dAy}
  For every $(X,d):\MS_Q$ with $Q$ continuous, $A\subseteq X$, $y\in X$,
  and $\delta\in Q$:
  \begin{equation*}
    \delta\ll d(A,y)\iff\exists A_0\subseteq_f A.
  \delta\ll d(A_0,y).
  \end{equation*}
  Moreover, if $Q$ is linear and $\qb\ne\delta$, then:
  \begin{equation*}
    \delta\ll d(A,y)\iff\exists x \in A.\delta\ll d(x,y).
  \end{equation*}
\end{lemma}

\begin{proof}
  Appendix~\ref{proof:lem:dAy}. \qed
\end{proof}

\begin{proposition}
  \label{prop:linear_quant_tdR_tdS}
  If $Q$ is a linear non-trivial quantale, then $\tau_{d,R}=\tau_{d,S}$.
  \end{proposition}

\begin{proof}
  Appendix~\ref{proof:prop:linear_quant_tdR_tdS}. \qed
\end{proof}

\begin{remark}
  Propositions~\ref{prop:linear_quant_tdR_tdS}~and~\ref{prop:spec-robust-top}
  ensure that, when $Q$ is linear and non-trivial, by applying the
  monad transformer $\T$ of Section~\ref{sec:Po-enriched}, we get a
  monad mapping a $Q$-metric space $(X,d)$ to the separated $Q$-metric
  space of closed subsets of $X$ with respect to the dual topology
  $\tau_d^o$ with the Hausdorff-Smyth metric. In this way, we recover
  the setting of \cite{MoggiFT-ICTCS-2019} as a special case.
\end{remark}

\begin{example}
  When the quantale $Q$ is not linear, the robust topology
  $\tau_{d,R}$ can be strictly finer than the *-robust topology
  $\tau_{d,S}$. For instance, consider the $Q$-metric space $(X,d)$,
  in which $Q = \RQ_+ \times \RQ_+$, $X= \RQ^2$, and the distance is
  given by $d((x,y),(x',y')) = (|x-x'|, |y-y'|)$. Let
  $\delta_0 \defeq (1, 1)$ and note that $\qI = (0,0)$. Take
  $A \defeq \setf{ (0,2), (2,0)} \subseteq \RQ^2$,
  $p \defeq (2,2) \in \RQ^2$, and consider the set
  $O \defeq \bigcup_{\delta_0\ll\delta'\ll\qI}
  \PS(B_R(A,\delta'))$. The set $O$ is in $\tau_{d,R}$, but it is not
  open in the *-robust topology $\tau_{d,S}$. The reason is that
  $d(A,p) = (0,0) = \qI$. Hence, for any $\delta \ll \qI$, the set
  $B_S(A,\delta)$ must contain $p$. But, the point $p$ is not included
  in any set in $O$, because
  $\forall p' \in A.\ (1, 1) \not \ll d(p',p)$.
\end{example}

\section{Concluding Remarks}
\label{sec:Concluding_remarks}

\paragraph{Related work.}
Flagg and Kopperman define $\V$-continuity spaces \cite[Def
3.1]{FlaggK97} and $\V$-domains, with $\V$ a \emph{value quantale}
\cite[Def 2.9]{FlaggK97}, \ie, the dual $\V^o$ of $\V$ is (in our
terminology) a commutative affine quantale, whose underlying complete
lattice is completely distributive---hence,
by~\cite[Thm.~7.1.1]{AbramskyJung94-DT}, continuous---and satisfies
additional properties formulated using a stronger variant $\lll$ of
the way-below relation $\ll$, called the \emph{totally-below}
relation, namely $p \lll q$ iff for any $A \subseteq Q$, if
$q \qle \qJ A$, then $\exists a \in A. p\qle a$ (in contrast with the
definition of $\ll$, the set $A$ is not required to be directed).
Thus, a $\V$-continuity space $(X,d)$ is what we call a $\V^o$-metric
space, while a $\V$-domain is a separated $\V^o$-metric space
satisfying further properties.
The metric $d_U$ in \cite[Sec 6]{FlaggK97} corresponds to our $d_S$,
and \cite[Thm 6.1]{FlaggK97} characterizes those $B$ such that
$d_U(A,B)=0$ as the subsets of the closure of $A$ in the topology
$\tau_d^o$, under the stronger assumption that $\V$ is a value
quantale.  The upper powerdomain $U(X)$ defined in \cite[Sec
  6]{FlaggK97} is almost the separated object equivalent to
$\PM_S(X)$, as its carrier consists of the closed subsets in the
topology $\tau_d^o$, except the empty one.

Although not every topology is induced by a classical metric,
Kopperman \cite{Kopperman:All_topologies_metric:1988} showed that all
topologies come from generalized metrics. Cook and
Weiss~\cite{CookW21} present a more nuanced discussion of this fact,
with constructions that avoid the shortcomings of Kopperman's original
construction. Their focus, however, is on comparing various topologies
that arise from a given generalized metric, \ie, those generated by
open sets, closed sets, interior, and exterior systems. Although the
four topologies coincide in classical metric spaces, they may be
different in quantale valued metric spaces. In particular, they
consider three conditions on a quantale, which are named Kuratowski,
Sierpi{\'n}ski, and triangularity
conditions~\cite[Def.~3]{CookW21}. When a commutative affine quantale
$Q$ satisfies these three conditions, it can be shown that all the
four topologies coincide for the metric spaces valued in $Q$.
Cook and Weiss~\cite{CookW21} use the totally-below relation $\lll$,
which is included in the way-below relation $\ll$.  Under the three
conditions they impose on quantales, one can show that for every
$\delta\ll\qI$ there exists $\delta'\lll\qI$ such that
$\delta\qle\delta'$.  Therefore, the topology generated by open balls
with radius $\delta'\lll\qI$ coincide with that generated by the open
balls with radius $\delta\ll\qI$.

The main drawback of value quantales and the quantales considered in
\cite{CookW21} is that they are not closed under product, which is
crucial for multi-dimensional quantitative analyses.
On the other hand, a continuous quantale $Q$ may not satisfy the
Kuratowski condition, and therefore the four topologies considered in
\cite{CookW21} for a $Q$-metric space may not coincide.  Specifically,
$d_S(A, \{x\}) = \qI$ may not entail that $x$ is in the closure of $A$
under the open ball topology.

\paragraph{Future work.}
The results of the current article may be regarded as the first steps
towards a framework for robustness analysis with respect to
perturbations that are measured using generalized metrics.  As such,
more remains to be done for development of the framework.  Our future
work will include study of effective structures on $Q$-metric spaces.

In~\cite{goubault2008simulation}, Goubault-Larrecq defines the
Hausdorff-Hoare and the Hutchinson hemi-metrics.  We plan to
investigate if they scale-up to $\Po$-enriched monads (or
endofunctors) on the category of $Q$-metric spaces, and in this case
apply to them the monad transformer defined in
Section~\ref{sec:Po-enriched}.

We also plan to study the impact of imprecision on probability
distributions on ($Q$-)metric spaces, \ie, to which extent they are
indistinguishable in the presence of imprecision, by applying our
monad transformer to probability monads.



\appendix

\section{Proofs}
\label{appendix:proofs}

\subsection{Proof of Lemma~\ref{lemma:inter:cq}}

\label{proof:lemma:inter:cq}

  We prove only the first implication.  If $q_1 \ll q_2$, then
\begin{eqnarray*}
q_1 \ll q_2 & = &q_2 \qT \qI \\  
(\text{by continuity of $Q$})& = & q_2 \qT \qJ\qwb\qI \\
(\text{by distributivity for $\qT$})& = & \qJ \setb{q_2 \qT q}{q\ll \qI}.
\end{eqnarray*}
Hence, for some $q \ll \qI$, we must have $q_1 \ll q_2 \qT q$, because
$\setb{q_2 \qT q}{q\ll \qI}$ is directed and $\setb{q'_2 \in Q}{q_1 \ll q'_2}$
is Scott open~\cite[Proposition~2.3.6]{AbramskyJung94-DT}.  \qed

\subsection{Proof of Proposition~\ref{prop:MetQ_has_small_sums_etc}}

\label{proof:prop:MetQ_has_small_sums_etc}

  Given a family $\famb{(X_i,d_i)}{i\in I}$ of objects in $\MS_Q$, the
  metric on the product $\Pi_{i \in I} X_i$ (computed in $\Set$) is
  $d_\Pi(x,y)=\qM_{i \in I}d_i(x_i,y_i)$, and the metric on the sum
  $\Sigma_{i \in I} X_i$ is $d_\Sigma((j,x),(j',x'))=d_j(x,x')$ if $j=j'$
  else $\bot$.

  Given a pair of short maps $f,g:\MS_Q((X,d),(X',d'))$, the equalizer
  is obtained by taking the equalizer $\iota : X_e \to X$ in
  $\Set$, \ie, $X_e=\setb{x \in X}{f(x)=g(x)}$ and $\iota$ is the
  inclusion of $X_e$ into $X$, and endowing $X_e$ with the
  restriction of $d$ to it. Then, $\iota$ is obviously short.
  Dually, the coequalizer is obtained by taking the coequalizer $\pi :
  X' \to X'/\approx$ in $\Set$, \ie, $\approx$ is the smallest
  equivalence relation on $X'$ including the relation
  $\setb{(f(x),g(x))}{x\in X}$ and $\pi$ is the quotient map $x\mapsto
  [x]$, and endowing $X'/\approx$ with the metric $d'_\approx$ given
  by $d'_\approx([x],[y]) =\qJ_{x'\in[x],y'\in[y]} d'(x',y')$.\qed
    

\subsection{Proof of Proposition~\ref{prop:ob:cq}}

\label{proof:prop:ob:cq}
  
For each property we give a proof hint.
\begin{enumerate}
\item follows from $\delta\ll \qI\qle d(x,x)$
\item follows from $\delta\qle\delta'\ll d(x,y)\implies\delta\ll d(x,y)$
\item
  $y\in B(x,\delta)$ is equivalent to $\delta\ll d(x,y)$, thus (by
  Lemma~\ref{lemma:inter:cq}) $\delta\ll d(x,y)\qT\delta'$ for some
  $\delta'\ll \qI$.
  Moreover, $B(y,\delta')\subseteq B(x,\delta)$ is equivalent to
  $\delta'\ll d(y,z)\implies \delta\ll d(x,z)$.
  If $\delta'\ll d(y,z)$, then $\delta\ll d(x,y)\qT\delta'\qle
  d(x,y)\qT d(y,z)\qle d(x,z)$, which implies (by
  Proposition~\ref{thm:wb:prop}) $\delta\ll d(x,z)$.

\item By item~\ref{ob:cq:3}, for $i \in \setf{1,2}$,
  $y\in B(x_i,\delta_i)$ implies
  $B(y,\delta'_i)\subseteq B(x,\delta_i)$ for some $\delta'_i\ll\qI$.
  Let $\delta'=\delta'_1\qj\delta'_2$, then
  $\delta'_i\qle\delta'\ll\qI$ (by
  Proposition~\ref{thm:wb:prop}). Thus,
  $B(y,\delta')\subseteq B(y,\delta'_i)\subseteq B(x,\delta_i)$ (by
  item~\ref{ob:cq:2}).\qed
\end{enumerate}

\subsection{Proof of Lemma~\ref{lemma:e-d:open}}
\label{proof:lemma:e-d:open}

  The $(\Leftarrow)$ direction is straightforward as
  property~\eqref{e-d:open} states that $O$ is a union of open
  balls. Thus, $O\in\tau_d$.
  To prove the $(\Rightarrow)$ direction, note that every $O\in\tau_d$
  is a union of finite intersections of open balls. Therefore, to
  prove that property~\eqref{e-d:open} holds for all $O\in\tau_d$,
  it suffices to prove that, for any $n \in \NQ$ and any (finite)
  sequence $\famb{B_i}{i \in n}$ of open balls,
  property~\eqref{e-d:open} holds for $\cap_{i \in n} B_i$, which we prove
  by induction on $n$:
  \begin{itemize}
  \item base case $0$: We note that $\cap \emptyset = X$ and
    $X=B(x,\bot)$ for any $x\in X$. Thus, by choosing $\delta = \bot$,
    $\cap \emptyset$ satisfies property~\eqref{e-d:open} by
    item~\ref{ob:cq:2} of Proposition~\ref{prop:ob:cq};
  
  \item inductive step $n+1$: by induction hypothesis $O=\cap_{i \in
    n} B_i$ satisfies property~(\ref{e-d:open}). Thus, for any $x\in
    O\cap B_n$, we have $B(x,\delta)\subseteq O$ for some
    $\delta\ll\qI$. In particular, $x\in B(x,\delta)\cap B_n$.
    Therefore, by item~\ref{ob:cq:4} of Proposition~\ref{prop:ob:cq},
    there exists $\delta'\ll\qI$ such that $B(x,\delta')\subseteq
    B(x,\delta)\cap B_n\subseteq O\cap B_n$. \qed
  \end{itemize}

  \subsection{Proof of Theorem~\ref{thm:e-d:continuity}}
\label{proof:thm:e-d:continuity}

  For the $(\Rightarrow)$ direction, assume that
  $f:\Top((X_1,\tau_{d_1}),(X_2,\tau_{d_2}))$, which means
  $\forall O\in\tau_{d_2}.\inv{f}(O)\in\tau_{d_1}$.  Let $O$ be the
  open ball $B(f(x),\epsilon)\in\tau_{d_2}$. Then, by
  Lemma~\ref{lemma:e-d:open}, there exists a $\delta\ll\qI_1$ such
  that $B(x,\delta)\subseteq\inv{f}(O)$, which is equivalent to
  $f(B(x,\delta))\subseteq O=B(f(x),\epsilon)$.
  
For the $(\Leftarrow)$ direction, assume that $O\in\tau_{d_2}$ and $f$
satisfies property~(\ref{e-d:continuity}). We must prove that
$O'=\inv{f}(O)\in\tau_{d_1}$, or equivalently (by
Proposition~\ref{prop:ob:cq}) for any $x\in O'$, there exists
$\delta\ll\qI$ such that $B(x,\delta)\subseteq O'$. If $x\in O'$, then
$f(x)\in O$. Hence, by Lemma~\ref{lemma:e-d:open},
$B(f(x),\epsilon)\subseteq O$ for some $\epsilon\ll\qI_2$.  By
property~(\ref{e-d:continuity}), there exists a $\delta\ll\qI_1$ such
that $f(B(x,\delta))\subseteq B(f(x),\epsilon)\subseteq O$, which
implies $B(x,\delta)\subseteq O'$.  \qed

\subsection{Proof of Proposition~\ref{prop:BR_A_delta_properties}}

\label{proof:prop:BR_A_delta_properties}

For each property we give a proof hint.
\begin{enumerate}
\item Follows easily from the definition of $B_R(A,\delta)$.
    
\item First, under the assumption $\delta_1,\delta_2\ll\qI$, one has
  $\delta_1\qT\delta_2\qle\delta_1\qT\qI=\delta_1\ll\qI$ and
  $\delta_1\qT\delta_2\qle\qI\qT\delta_2=\delta_2\ll\qI$.
  If $z\in B_R(B_R(A,\delta_1),\delta_2)$, then $\delta_2\ll
  d(y,z)$ for some $y\in B_R(A,\delta_1)$, thus $\delta_2\ll d(y,z)$
  and $\delta_1\ll d(x,y)$ for some $x\in A$, thus
  $\delta\ll\delta_1\qT\delta_2\qle d(x,y)\qT d(y,z)\qle d(x,z)$.

\item Follows easily from Lemma~\ref{lemma:e-d:closed} and the
  definition of $B_R(A,\delta)$.
  
\item It suffices to prove the inclusion
  $B_R(\cl{A}^o,\delta)\subseteq B_R(A,\delta)$.  If $z\in
  B_R(\cl{A}^o,\delta)$, then $\delta\ll d(y,z)$ for some
  $y\in\cl{A}^o$.  Choose $\epsilon\ll\qI$ such that
  $\delta\ll\epsilon\qT d(y,z)$ and $x\in A$ such that $\epsilon\ll
  d(x,y)$, then $\delta\ll\epsilon\qT d(y,z)\qle d(x,y)\qT d(y,z)\qle
  d(x,z)$.

\item The first inclusion follows easily from the definition of
  $A_\delta$, For the second inclusion, since $A_\delta\subseteq
  \cap_{\epsilon\ll\qI}B_R(B_R(A,\delta),\epsilon)$, it suffices to choose
  $\epsilon\ll\qI$ such that $\delta'\ll\delta\qT\epsilon$, then
  $B_R(B_R(A,\delta),\epsilon)\subseteq B_R(A,\delta')$.
  \qed
\end{enumerate}

\subsection{Proof of Proposition~\ref{prop:spec-robust-top}}
\label{proof:prop:spec-robust-top}

For the $(\Leftarrow)$ direction, consider $U \in \tau_{d,R}$ such
that $A\in U$.  By definition of $\tau_{d,R}$, we have
$\PS(B_R(A,\delta)) \subseteq U$, for some $\delta \ll \qI$.  By
hypothesis and Proposition~\ref{prop:BR_A_delta_properties} we get
$B \subseteq \cl{A}^o \subseteq B_R(\cl{A}^o,\delta) =
B_R(A,\delta)$. Thus, $B \in U$, as required.

For the $(\Rightarrow)$ direction, we proceed by contraposition,
namely, we prove the logically equivalent $B \not\subseteq \cl{A}^o
\implies(\exists U\in\tau_{d,R}.A\in U\land B\notin U)$.
If $B \not\subseteq \cl{A}^o$, then there is $x\in B$ such that
$x\notin \cl{A}^o$.
By Proposition~\ref{prop:BR_A_delta_properties}, we have $\cl{A}^o
= \bigcap_{\delta\ll\qI} B_R(A,\delta)$. Thus, there is
$\delta \ll \qI$ such that $x \notin B_R(A,\delta)$, and consequently,
$x\notin B_R(A,\delta')$ for every $\delta'$ such that
$\delta\ll \delta'\ll \qI$.
We define an open subset $U\in\tau_{d,R}$ such that $A\in U$ and
$B\not\in U$.  Let
$U=\bigcup_{\delta\ll\delta'\ll\qI} \PS(B_R(A,\delta'))$. Clearly,
$A\in U$, because by Lemma~\ref{lemma:inter:cl} there is at least one
$\delta'$ such that $\delta\ll \delta'\ll \qI$, and $B\not\in U$,
since $x\notin B_R(A,\delta')$ for every $\delta'$ such that
$\delta\ll \delta'\ll \qI$.

It remains to prove that $U\in\tau_{d,R}$, namely, that for
every $A' \in U$, \ie, $A'\subseteq B_R(A,\delta_1)$ for some
$\delta\ll\delta_1\ll\qI$, there exists $\delta_2\ll\qI$ such that
$\PS(B_R(A',\delta_2))\subseteq U$.
By Lemma~\ref{lemma:inter:cl}~and~\ref{lemma:inter:cq}, there are
$\delta',\delta_2\ll\qI$ such that
$\delta \ll \delta' \ll \delta_1\qT\delta_2 \ll \qI$.
Hence, by Proposition~\ref{prop:BR_A_delta_properties} we get
$B_R(A',\delta_2) \subseteq B_R(B_R(A,\delta_1),\delta_2) \subseteq
B_R(A,\delta') $.  Therefore, we have
$\PS(B_R(A',\delta_2)) \subseteq \PS(B_R(A,\delta')) \subseteq U$.
\qed

\subsection{Proof of
  Proposition~\ref{prop:MetQ_every_object_separated}}

\label{proof:prop:MetQ_every_object_separated}

Given a $Q$-metric space $(X,d)$, denote by $\sim_d$ the equivalence
induced by the preorder $\leq_d$ on $X$, \ie, $x\sim_d y\defiff\qI\qle
d(x,y)\land \qI\qle d(y,x)$.
Let $X_0$ be the quotient $X/\sim_d$ and define $d_0:X_0\times X_0 \to
Q$ as $d_0(\eqc{x},\eqc{y}) = d(x,y)$.  Since $x\sim_d x'\land y\sim_d
y'\implies d(x,y)=d(x',y')$, $d_0$ is a well-defined $Q$-metric on
$X_0$.

Let $\r:X \to X_0$ be the map such that $\r(x)=\eqc{x}$, which is
clearly an isometry from $(X,d)$ to $(X_0,d_0)$.
Since $\r$ is surjective, there is a section $\sr:X_0 \to X$, which
chooses a representative from each equivalence class $\eqc{x}\in
X_0$. Thus, $\sr(\eqc{x} )\sim_d x$ for every $x\in X$.
Therefore, $\sr$ in an isometry from $(X_0,d_0)$ to $(X,d)$.

To prove that $(\r,\sr)$ is an equivalence in $\MS_Q$, \ie,
$\r\circ\sr\sim\id_{X_0}$ and $\sr\circ\r\sim\id_{X}$, where $\sim$ on
$\MS_Q((X,d),(X',d'))$ is the pointwise extension of $\sim_{d'}$, it
suffices to observe that $\r(\sr(\eqc{x}))=\eqc{x}$ and
$\sr(\r(x))\sim x$ for every $x\in X$.  \qed

\subsection{Proof of Theorem~\ref{thm:endo:transformer}}

\label{proof:thm:endo:transformer}

  All the maps on hom-preorders used in the definition of $\T$ are
  monotonic, thus they preserve $\sim$.  Therefore, to prove that two
  arrows $f,g\in\A(X,Y)$ defined by different monotonic constructions
  are equal, it suffices to prove that they are equivalent (\ie,
  $f\sim g$), if $Y$ is separated.
  For the same reason, if in a monotonic construction, one can replace 
  $\sr_X$ with another $\sr'_X$ such that $(\r_X,\sr'_X)$ is an
  equivalence, the results will be equivalent, because
  $\sr_X\sim\sr'_X$.

  \begin{itemize} 
  \item $\T\hat{M}=(M',\eta',-^{*'})$ satisfies 
    equations~(\ref{eq:monad}) for a monad, namely:
    \begin{itemize}
    \item $(\eta'_X)^{*'}=\id_{M'X}:\R(MX)\to\R(MX)$, because:
      \begin{align*} 
        {\eta'_X}^{*'} 
          &= \r_{MX}\circ(\sr_{MX}\circ\r_{MX}\circ\eta_X)^*\circ\sr_{MX} \\ 
          &\sim \r_{MX}\circ\eta_X^*\circ\sr_{MX} 
           = \r_{MX}\circ\sr_{MX} 
           = \id_{M'X}.
       \end{align*} 
    \item $f^{*'}\circ\eta'_X=f:X\to\R(MY)$ when $f:X\to\R(MY)$,
      because:
      \begin{align*}    
        f^{*'}\circ\eta'_X 
          &= \r_{MY}\circ(\sr_{MY}\circ f)^*\circ\sr_{MX}\circ\r_{MX}\circ\eta_X \\ 
          &\sim \r_{MY}\circ(\sr_{MY}\circ f)^*\circ\eta_X
           = \r_{MY}\circ\sr_{MY}\circ f 
           = f .
      \end{align*} 
    \item $g^{*'}\circ f^{*'}=(g^{*'}\circ f)^{*'}:\R(MX)\to\R(MZ)$
      when $f:X\to\R(MY)$ and $g:Y\to\R(MZ)$, because:
      \begin{align*} 
        g^{*'}\circ f^{*'}
          &= \r_{MZ}\circ(\sr_{MZ}\circ g)^*\circ\sr_{MY}\circ
             \r_{MY}\circ(\sr_{MY}\circ f)^*\circ\sr_{MX} \\ 
          &\sim \r_{MZ}\circ(\sr_{MZ}\circ g)^*\circ(\sr_{MY}\circ f)^*\circ\sr_{MX} \\ 
          &= \r_{MZ}\circ((\sr_{MZ}\circ g)^*\circ\sr_{MY}\circ f)^*\circ\sr_{MX} \\ 
          &\sim \r_{MZ}\circ(\sr_{MZ}\circ\r_{MZ}\circ
              (\sr_{MZ}\circ g)^*\circ\sr_{MY}\circ f)^*\circ\sr_{MX} \\ 
          &= \r_{MZ}\circ(\sr_{MZ}\circ g^{*'}\circ f)^*\circ\sr_{MX} 
           = (g^{*'}\circ f)^{*'}  .
      \end{align*}
    \end{itemize}
  \item $\inT_{\hat{M}}$ satisfies 
    equations~(\ref{eq:monad:map}) for a monad map from $\hat{M}$ to
    $\T\hat{M}$, namely:
    \begin{itemize}
    \item $\inT_{\hat{M},X}\circ\eta_X=\eta'_X:X\to\R(MX)$, because:
      \begin{align*} 
        \inT_{\hat{M},X}\circ\eta_X
          &= \r_{MX}\circ\eta_X
           = \eta'_X .
      \end{align*} 
    \item $\inT_{\hat{M},Y}\circ f^*=(\inT_{\hat{M},Y}\circ
      f)^{*'}\circ\inT_{\hat{M},X}:MX\to\R(MY)$ when $f:X\to MY$, because:
      \begin{align*} 
        \inT_{\hat{M},Y}\circ f^*
          &= \r_{MY}\circ f^* \\ 
          &\sim \r_{MY}\circ(\sr_{MY}\circ\r_{MY}\circ f)^*\circ\sr_{MX}\circ\r_{MX} \\ 
          &= (\inT_{\hat{M},Y}\circ f)^{*'}\circ\inT_{\hat{M},X} .
      \end{align*} 
  \end{itemize}
  \end{itemize}
  \qed

\subsection{Proof of Proposition~\ref{prop:PS_monad_MetQ}}

\label{proof:prop:PS_monad_MetQ}
  
We prove each of the four properties in sequence.
\begin{enumerate}
\item $\qI\qle d_S(A,A)$ means
  $\forall y \in A.\qI\qle\qJ_{x \in A}d(x,y)$. It holds because
  $\qI\qle d(y,y)\qle\qJ_{x \in A}d(x,y)$ for any $y \in A$. The
  inequality $d_S(A,B)\qT d_S(B,C)\qle d_S(A,C)$ is equivalent to
  $\forall z \in C.d_S(A,B)\qT d_S(B,C)\qle\qJ_{x \in A}d(x,z)$, which holds
  by the following chain of $\qle$ for any $z \in C$
  \begin{eqnarray*}
  d_S(A,B)\qT d_S(B,C)&\qle&\text{by monotonicity of $\qT$}\\
  d_S(A,B)\qT\qJ_{y \in B}d(y,z)&=&\text{by distributivity}\\
  \qJ_{y \in B}(d_S(A,B)\qT d(y,z))&\qle&\text{by monotonicity of $\qT$}\\
  \qJ_{y \in B}(\qJ_{x \in A}d(x,y))\qT d(y,z)&=&\text{by distributivity}\\
  \qJ_{y \in B}\qJ_{x \in A}(d(x,y)\qT d(y,z))&\qle&\text{by triangular inequality}\\
  \qJ_{y \in B}(\qJ_{x \in A}d(x,z))&\qle&\text{because $\qJ_{j \in J}q\qle q$}\\
  \qJ_{x \in A}d(x,z).
  \end{eqnarray*}

\item The property follows from $d_S(\{x\},\{x\})=d(x,x)$. Actually
  $\eta$ is an isometry.

\item The implication amounts to proving
  $d_S(A,B)\qle d'_S(f^*(A),f^*(B))$ from the assumption
  $\forall x,y \in X.d(x,y)\qle d'_S(f(x),f(y))$.
  But:
  \begin{equation*}
  d_S(A,B)\qle d'_S(f^*(A),f^*(B))  
  \end{equation*}
 means
  $\forall y \in B.\forall y' \in f(y).d_S(A,B)\qle\qJ_{x' \in
    f^*(A)}d'(x',y')$. Thus, it holds by the following chain of $\qle$
  for $y \in B$ and $y' \in f(y)$:
  \begin{eqnarray*}
  d_S(A,B)&\qle&\text{because $\forall k \in J.(\qM_{j \in J}q_j)\qle q_k$}\\
  \qJ_{x \in A}d(x,y)&\qle&\text{by the assumption}\\
  \qJ_{x \in A}d'_S(f(x),f(y))&\qle&\text{because $\forall k \in J.(\qM_{j \in J}q_j)\qle q_k$}\\
  \qJ_{x \in A}(\qJ_{x' \in f(x)}d'(x',y'))&=&\text{by definition of $f^*$}\\
  \qJ_{x' \in f^*(A)}d'(x',y').
  \end{eqnarray*}

\item The implication amounts to proving
  $\forall A \in \PM(X).\qI\qle d'_S(f^*(A),g^*(A))$ from the
  assumption $\forall y \in X.\qI\qle d'_S(f(y),g(y))$, \ie,
  $\forall y \in A.\forall y' \in g(y).\qI\qle\qJ_{x' \in
    f(y)}d'(x',y')$.
  But $\qI\qle d'_S(f^*(A),g^*(A))$ means
  $\forall y \in A.\forall y' \in g(y).\qI\qle\qJ_{x' \in
    f^*(A)}d'(x',y')$. Thus, it holds by the following chain of $\qle$
  for any $y \in A$ and $y' \in g(y)$:
  \begin{eqnarray*}
  \qI&\qle&\text{by the assumption}\\
  \qJ_{x' \in f(y)}d'(x',y')&=&\text{by definition of $f^*$}\\
  \qJ_{x' \in f^*(A)}d'(x',y').
  \end{eqnarray*}
\end{enumerate}
Since the unit $\eta$ and Kleisli extension $-^*$ for $\PM_S$ are
\emph{equal} to those for the monad $\PM$ on $\Set$, they necessarily
satisfy the required equational properties.  \qed

\subsection{Proof of Lemma~\ref{lemma:BS_delta_properties}}

\label{proof:lemma:BS_delta_properties}

For each property we give a proof hint.
\begin{enumerate}
\item The first inclusion follows from $d(x,y)\qle d(A,y)$ for every
  $x\in A$, while the second follows from $d(A,y)\qle d(A',y)$ when
  $A\subseteq A'$.

\item Let $B=B_S(A,\delta)$. Then, $\forall y \in B.\delta\ll d(A,y)$. Thus:
\begin{equation*}
\delta\qle\qM_{y \in B}\delta\qle\qM_{y \in B}d(A,y)=d_S(A,B).
\end{equation*}
  \qed
\end{enumerate}

\subsection{Proof of Proposition~\ref{prop:top:SR}}
\label{proof:prop:top:SR}

  $\tau_{d,S}\subseteq\tau_{d,R}$ follows from $B_R(A,\delta)\subseteq
  B_S(A,\delta)$ when $A\subseteq X$ and $\delta\ll\qI$.
  
  Let $B(A,\delta)$ be the open ball with center $A\subseteq X$ and radius
  $\delta\ll\qI$ for the metric $d_S$.
  To prove $\tau_{d_S}\subseteq\tau_{d,S}$, we show that every open
  ball $B(A,\delta)$ belongs to $\tau_{d,S}$.
  Since $B(A,\delta)$ is downwards closed, it suffices to prove that
  $B\in B(A,\delta)\implies \exists\epsilon\ll\qI.B_S(B,\epsilon)\in
  B(A,\delta)$.
  Choose $\epsilon\ll\qI$ such that $\delta\ll d_S(A,B)\qT\epsilon$,
  then $\delta\ll d_S(A,B)\qT\epsilon\qle d_S(A,B)\qT
  d_S(B,B_S(B,\epsilon))\qle d_S(A,B_S(B,\epsilon))$.
  To prove $\tau_{d,S}\subseteq\tau_{d_S}$, we show that
  $B(A,\delta)\subseteq\PS(B_S(A,\delta))$ for every $A\subseteq X$ and
  $\delta\ll\qI$.  In fact, $d_S(A,B)=\qM_{y \in B}d(A,y)\qle d(A,y)$ when
  $B\subseteq X$ and $y\in B$.  \qed

\subsection{Proof of Lemma~\ref{lem:dAy}}
\label{proof:lem:dAy}
Choose (using Lemma~\ref{lemma:inter:cl}) $\delta'$ such that
$\delta\ll\delta'\ll d(A,y)$, then we have the chain of
equivalences: \begin{itemize} \item $\delta\ll d(A,y)\iff$ by
definition of $d(A,y)$
  
  \item $\delta\ll\delta'\ll\qJ_{x \in A}d(x,y)\iff$ by definition of $\ll$
  
  \item $\exists A_0\subseteq_f A.
    \delta\ll\delta'\qle\qJ_{x \in A_0}d(x,y)=d(A_0,y)$.
  \end{itemize}
  If $Q$ is linear and $\qb\qlt\delta$, then $\qb\qlt d(A_0,y)$, thus
  $\emptyset\subset A_0\subseteq_f A$.  This implies that
  $\setb{d(x,y)}{x\in A_0}$ has a maximum, thus
  $d(A_0,y)=d(x,y)$ for some $x\in A_0$.\qed

\subsection{Proof of Proposition~\ref{prop:linear_quant_tdR_tdS}}

\label{proof:prop:linear_quant_tdR_tdS}

  $\tau_{d,S}\subseteq\tau_{d,R}$ follows from
  Proposition~\ref{prop:top:SR}.
  For the other inclusion we prove that
  $\forall\delta\ll\qI.\exists\epsilon\ll\qI.B_S(A,\epsilon)\subseteq
  B_R(A,\delta)$.  By Lemma~\ref{lem:dAy},
  $B_S(A,\delta)=B_R(A,\delta)$ when $\qb\qlt\delta$.
  $B_S(A,\qb)=X=B_R(A,\qb)$ when $\emptyset\subset A$.
  $B_R(\emptyset,\qb)=\emptyset=B_S(\emptyset,\delta)$ for any
  $\delta\ll\qI$ such that $\qb\qlt\delta$, which exists because $Q$
  is not trivial.\qed


\end{document}